\documentclass[fleqn,twoside]{article}
\usepackage{espcrc2}

\usepackage{graphicx,amssymb}

\title{What is the trouble with Dyson--Schwinger equations?}

\author{D.\ Kreimer\address{Institut des Hautes \'Etudes Scientifiques, \\
        35, rte.\ de Chartres, F-91440 Bures-sur-Yvette, France}%
        \thanks{Supported by CNRS. Work partially supported by NSF grant DMS-0401262; Center for Math.\ Phys., Boston University; BU-CMP/04-05.}}

\begin{document}

\def\R{\mathbb{R}}
\def\C{\mathbb{C}}
\def\N{\mathbb{N}}
\def\Q{\mathbb{Q}}
\def\Z{\mathbb{Z}}
\def\S{\mathbb{S}}
\def\s{\mathcal{S}}
\def\i{\mathcal{I}}
\def\T{\mathbb{T}}
\def\One{\mathbb{I}}
\def\F{\mathbb{F}}
\def\D{\mathcal{D}}
\def\Lhi{\mathcal{L}}
\def\W{\mathcal{W}}
\def\eps{\varepsilon}

\newcommand{\be}{\begin{equation}}
\newcommand{\ee}{\end{equation}}
\newcommand{\bea}{\begin{eqnarray}}
\newcommand{\eea}{\end{eqnarray}}
\newcommand{\beas}{\begin{eqnarray*}}
\newcommand{\eeas}{\end{eqnarray*}}

\newtheorem{theorem}{Theorem}
\newtheorem{lemma}[theorem]{Lemma}
\newtheorem{prop}[theorem]{Proposition}
\newtheorem{cor}[theorem]{Corollary}
\newtheorem{defn}[theorem]{Definition}
\newtheorem{rem}[theorem]{Remark}
\newtheorem{conj}[theorem]{Conjecture}
\newtheorem{exam}[theorem]{Example}

\begin{abstract}
We discuss similarities and differences between Green Functions in Quantum Field Theory and polylogarithms. Both can be obtained as solutions of fixpoint
equations which originate from an underlying Hopf algebra structure. Typically, the equation is linear for the polylog, and non-linear for Green Functions. We
argue though that the crucial difference lies not in the non-linearity of the latter, but in the appearance of non-trivial representation theory related to
transcendental extensions of the number field which governs the linear solution. An example is studied to illuminate this point. \vspace{1pc}
\end{abstract}

\maketitle

There is a close connection between the polylog and Green functions in quantum field theory \cite{DK1}. Indeed, the polylog can be regarded as a solution to a
linear Dyson--Schwinger equation (DSE), while Green Functions provide solutions to non-linear DSEs. For the polylog, consider, for suitable $z$ off the cut, \be
Li(\alpha,z)=1-\frac{1}{1-z}+\alpha\int_0^z\frac{Li(\alpha,x)}{x}dx.\ee This determines $Li(\alpha;z)=-\sum_{j=0}^\infty {\rm Li}_j(z)\alpha^j$ as the generating
function for the polylog ${\rm Li}_j(z)=\sum_n z^n/n^j$.

For the polylog and also for QFT Green functions, one finds these equations by first defining a Hopf algebra structure. Typically, this is a commutative and
connected graded Hopf algebra. One then studies its Hochschild cohomology, and identifies its closed one-cocycles, which are maps $B_+: H\to H$ such that
\begin{equation}bB_+=0 \Leftrightarrow \Delta B_+=B_+\otimes e +({\rm id}\otimes B_+)\Delta. \label{B_+}\end{equation}
For the polylog above, it suffices to consider the simplest of such Hopf algebras, the free commutative algebra on generators $t_n$, $n\in\N$, and declaring the
map $B_+:\; H\to H,$ $ B_+(t_n)=t_{n+1}$ to be a closed Hochschild one-cocycle so that we get a coproduct $\Delta(t_n)=\sum_{j=0}^n t_j\otimes t_{n-j}$ (we
identify $t_0$ with the unit in the Hopf algebra). This leads immediately to a distinguished combinatorial equation which determines the above generating
function (Green function) for the polylog: \be X=e+\alpha B_+(X)\;\Rightarrow X=\sum_{k\geq 0}\alpha^k t_k.\ee Upon introducing suitable Feynman rules $\phi$
which map $B_+$ to an integral operator one then reobtains the above integral equation as the image under the Feynman rules of this combinatorial equation
\cite{DK1}.

In general, to turn such a combinatorial fixpoint equation into an integral or differential equation, it suffices to define Feynman rules which effectively map
the Hochschild one-cocycles into integral operators. For the polylogs, this leads to their familiar iterated integrals, while it leads
 to the Feynman rules for perturbative
QFT, upon expanding the resulting integral equations, which are just the Dyson--Schwinger equations in their standard form. Note that this allows to derive the
Feynman rules and Dyson--Schwinger equations in a completely rigorous manner: the input needed is free quantum field theory, and the Hochschild cohomology of the
Hopf algebra which comes with it, once its free propagators and local interactions are chosen. The Hochschild cohomology of renormalizable field theories is
fairly well understood, the question is if efficient methods can be developed to come to conclusions concerning the existence and nature of solutions to DSEs.

While for the polylogarithm, a linear DSE as above suffices to define the generating function $Li(\alpha;z)$, Green functions in
quantum field theory generalize this situation by the choice of more complicated Hopf algebras, and hence the existence of more and more complicated one-cocycles
$B_+^p$, typically parametrized by primitive elements $p$ of the Hopf algebra of graphs \cite{annals}.

The identification of these one cocycles always leads to a combinatorial Dyson--Schwinger equation: \bea \Gamma^{\underline{r}} & = &  1 + \sum_{{p\in H_L^{[1]}}
\atop {{\rm res}(p)
  =  \underline{r}}} \frac{\alpha^{\vert p\vert}}{{\rm Sym}(p)}
B_+^p(X_{p})\nonumber\\ & = & 1+\sum_{{\Gamma\in H_L}\atop{ {\rm res}(\Gamma)=\underline{r}}}\frac{\alpha^{\vert\Gamma\vert}\Gamma}{{\rm
Sym}(\Gamma)}\;,\label{DSE}\eea
 where the first sum is over a finite (or
countable) set of Hopf algebra primitives $p$, \be \Delta(p)=p\otimes e + e \otimes p,\ee  indexing the closed Hochschild 1-cocycles $B_+^{p}$ above, while the
second sum is over all one-particle irreducible graphs contributing to the desired Green function, all weighted by their symmetry factors. Here, $X_{p}$ is a
polynomial in all $\Gamma^{\underline{r}}$, and the superscript $\underline{r}$ ranges over the finite set (in a renormalizable theory) of superficially
divergent Green functions. It indicates the number and type of external legs. We use ${\rm res}(p)=\underline{r}$ to indicate that the external legs of the graph
$p$ are of type $\underline{r}$. The structure of these equations allows for a proof of locality using Hochschild cohomology \cite{annals}, which is also evident
using a coordinate space approach \cite{Bergbau}.

These fixpoint equations are solved by an Ansatz
\begin{equation} \Gamma^{\underline{r}}=1+\sum_{k=1}^\infty \alpha^k c^{\underline{r}}_k .\end{equation}
We grant ourselves the freedom to call such an equation a DSE or a combinatorial equation of motion for a simple reason: the DSEs of any renormalizable quantum
field theory can be cast into this form. Crucially, in the above it can be shown that \be X_{p}=\Gamma^{{\rm res}(p)}X_{\rm coupl}^{\mid p \mid},\ee where
$X_{\rm coupl}$ is a connected Green functions which maps to an invariant charge under the Feynman rules. This is rather obvious: consider, as an example, the
vertex function in quantum electrodynamics: a $n$ loop primitive graph $p$ contributing to it provides $2n+1$ internal vertices, $2n$ internal fermion
propagators and $n$ internal photon propagators. An invariant charge \cite{Gross} is provided by a vertex function multiplied by the squareroot of the photon
propagator and the fermion propagator. Thus the integral kernel corresponding to $p$ is dressed by $2n$ invariant charges, and one vertex function. This is a
general fact: each integral kernel corresponding to a Green function with external legs $\underline{r}$ in a renormalizable quantum field theory is dressed by a
suitable power of invariant charges proportional to the grading of that kernel, and one additional apperance of $\Gamma^{\underline{r}}$ itself. This immediately
shows that for a vanishing $\beta$-function the DSEs are reduced to a linear set of equations, and that the general case can be most efficiently handled by an
expansion in the breaking of conformal symmetry induced by a non-vanishing $\beta$-function. Thus, a complete understanding of the linear case goes a long way in
understanding the full solution. This emphasizes the crucial role which the insertion-elimination Lie algebra \cite{InsElim} in the ladder case \cite{Menc} plays
in the full theory.

The difference between a vanishing and a non-vanishing $\beta$-function expresses itself in three remarkable ways: for a vanishing $\beta$-function, the solution
to a Dyson--Schwinger equation provides for a cocommutative Hopf algebra, it provides a group-like element in the Hopf algebra, and its coefficients evaluated by
the Feynman rules provide polylogs which are transcendentally extended for a non-vanishing $\beta$-function, in comparison with the accompanying solution of the
linear case.

For the case of a linear DSE one immediately sees that Fourier transform allows a solution once the Hochschild closed one-cocycles have been determined, as
exhibited below. For the non-linear case, the propagator-coupling duality of \cite{BK} gave a first sign that the Hopf algebra structure provides sufficient
structure to solve a DSE even in the non-linear case. Indeed, the highly non-linear DSE \be X=1-\alpha B_+^p\left(\frac{1}{X}\right)\ee in the Hopf algebra of
undecorated rooted trees has a solution for any Feynman rule which maps $B_+^p$ into an integral operator for a primitively divergent Feynman integral
corresponding to $p$, see \cite{BK} for examples. This equation is already cast into the form of Eq.(\ref{DSE}) above, and thus determines \be X_{\rm
coupl}=\frac{1}{[X]^2}.\ee This explains the structure of the resulting propagator coupling duality of \cite{BK}\be \frac{\partial \log G(\alpha,L)}{\partial
L}=\gamma\left(\frac{\alpha}{[G(\alpha,L)]^2}\right),\ee where $L=\log q^2/\mu^2$ and \be\gamma(\alpha)=\left[\partial \log G(\alpha,L)/\partial
L\right]_{L=0}\ee is the anomalous dimension. This can indeed be confirmed by Fourier analysis due to the fact that the non-linear corrections in that case can
be handled via derivatives on the transform of the corresponding integral operator associated to $p$, as there was only a single loop in the underlying primitive
$p$: \bea
& & \int d^4k \log(k^2/\mu^2)\left(\frac{k^2}{\mu^2}\right)^{-\gamma}K_p(k,q)\nonumber\\
 & & = -\partial_{\gamma}\int d^4k \left(\frac{k^2}{\mu^2}\right)^{-\gamma}K_p(k,q)\nonumber\\  & & =  -\partial_\gamma
T[K_p](\gamma),\label{der} \eea for an integral  kernel $K_p$ with Fourier transform $T[K_p](\gamma)$ associated to the primitive $p$.

In general, a hard but very interesting problem remains: if the $\beta$-function is non-va\-nishing, and the invariant charges thus not constant, the logarithmic
corrections in the integral kernel can not be handled as such derivatives due to the fact that these corrections appear at various different places, and thus
modify different propagators, while the Fourier transform for such Feynman integral kernels is defined via its overall momentum. Only for one-loop kernels do
they necessarily match. This leads, if one arranges for more than a single insertion place, naturally to the appearance of transcendental extensions, as the
study below shows.

We will hence emphasize the difference betwen two situations. We will first consider a linear DSE for a Green function in QFT, summarizing standard results
\cite{DK1}, showing how to solve it using Fourier analysis. We then argue that the propagator coupling duality of \cite{BK} exhibited above does not
suffice to handle the non-linear problem in general: the appearance of unknown transcendental extensions of the underlying field of transcendental numbers has to
be understood. We then study an example to show how these transcendental extensions relate to an underlying symmetry in our equation, comparable to the situation
in the study of finite algebraic extension in algebraic number theory, only that we have transcendental extensions. The point we want to make is that it is not
the non-linearity, but the appearance of these symmetries and transcendental extensions in the equations of motions which renders them difficult.

As a concrete example to discuss the linear case, we consider a three-point vertex function $\Gamma(q,\alpha)$. We let $K(l,q;\alpha)$ be a connected Green
function, the kernel, so that
\begin{equation}
\Gamma(q;\alpha)=1+\int d^4l \Gamma(l;\alpha)K(l,q;\alpha).\label{exa}
\end{equation}
is our DSE.  We assume that $K(l,q;\alpha)$ has a perturbative expansion $K=\sum_{p} \alpha^{\mid p \mid}K_p(l,q)$, where $K_p$ are integral kernels for
logaritmically divergent primitive vertex graphs constructed from
 the
sum of all overall convergent subdivergence-free 1PI graphs which are 2PI in the forward scattering channel. The above provides a sum of
logarithmically divergent integrals. Nevertheless, this DSE can be solved without recourse to regularization or renormalization making use of Fourier analysis on
the multiplicative group $\mathbb{R}^+$ to be introduced in a moment. What is required though is, as usual, a fixing of a boundary condition:
$\Gamma(q;\alpha)_{q^2=1}=1$.

Before we discuss this point, let us note that this integral equation has a Hopf algebraic backbone:
let $p$ be a graph appearing in the above kernel expansion, and consider the equation
\begin{equation}
X=e+\sum_{p} \alpha^{\mid p\mid}B_+^{p}(X).\label{formal}
\end{equation}
This equation defines a formal series in the Hopf algebra of words made out of letters $p$ whose solution is a combinatorial Euler product based on a
normalized shuffle
product \cite{annals}
\begin{equation} X=\prod_p^\vee \frac{1}{1-\alpha^{\mid p\mid}p}.\label{fact}\end{equation} If we define Feynman rules $\phi$ by $\phi(e)=1$ and
\begin{equation}
\phi(B_+^{p}(W))=\int d^4l K_p(l,q)\phi(W)(l),
\end{equation}
then $\phi(X)$ fulfills the same equation as $\Gamma$ in Eq.(\ref{exa}) above iff $X$ fulfills Eq.(\ref{formal}).
Furthermore, the maps $B_+^p$ are Hochschild closed in the sense of Eq.(\ref{B_+}) above.

Let us solve this equation by Fourier analysis with respect to the multiplicative group $\mathbb{R}^+$. We note that the closed Hochschild one-cocycle $B_+^p$
maps to an
integral operator for the kernel $K_p$. This integral operator is obtained by applying the Feynman rules to the underlying primitive graph $p$.
If we integrate out all
internal momenta of the four-point kernel, and also the angular integration of the final loop momentum, we are left with an integral of the form
\begin{equation}
I_p(q)=\int_0^\infty  K_p(r;q) d\log r,
\end{equation}
where $K_p(r;q)$ vanishes for $r=0$ and tends towards a constant $r_p$ (the residue of that graph $p$) for $r\to\infty$. Let us set $q^2=1$. We then define the
transform $T[K_p](\gamma)$ to be the integral \bea T[K_p](\gamma) & = & \int_0^\infty  \exp(-\gamma\log r)\times \nonumber\\
 & & \times K_p(r;q)\mid_{q^2=1} d\log r.\eea
Thus the transform is the Fourier transform wrt to the multiplicative group of the positive reals \cite{Edwards} of the integral kernel provided by our
underlying skeleton graph. Our Fourier analysis leads to an implicit equation:
\begin{equation}
1=\sum_p \alpha^{\mid p \mid} T[K_p](\gamma).
\end{equation}
This determines $\gamma=\gamma(\alpha)$ as a function of the coupling,
or vice versa the latter as a function $\alpha=\alpha(\gamma)$ of the anomalous dimension, standard practice in quantum field theory if one can ignore
the anomalies introduced by the $\beta$-function \cite{Delb}.

Consider as an example the Feynman integral $p_1$ corresponding to the one-loop vertex graph  in $\phi^3_6$ and its transform: \bea T[K_{p_1}] & = & \int d^6k
[k^2]^{-\gamma} \frac{1}{k^4(k+q)^2}\mid_{q^2=1}\nonumber\\ & = & \frac{1}{\gamma(1-\gamma)(1+\gamma)(2-\gamma)}. \eea Hence, we get
\begin{equation}
1=\alpha\frac{1}{\gamma(1-\gamma)(1+\gamma)(2-\gamma)},
\end{equation}
which determines the solution of Eq.(\ref{exa}) with $K=K_{p_1}$ as $\Gamma(q;\alpha)=[q^2]^{-\gamma(\alpha)}$, where
\be\gamma(\alpha)=1-\sqrt{5-4\sqrt{1-\alpha}}.\ee
The perturbative expansion is then easily regained by Taylor expanding this solution in
$\alpha$. The higher coefficients of $\gamma$ in $\alpha$ correspond precisely to the residues
\begin{equation}
L^{(p_1)}_m:=\oint d\rho [-\partial_\rho]^{m-1}\int d^6k  \frac{[k^2]^{-\rho}}{k^4(k+q)^2},
\end{equation}
the case $m=1$ defines the residue $r_p\equiv L_1^{p_1}$ above. In passing, we remark that this allows to factorize the result reflecting the combinatorial
factorization Eq.(\ref{fact}) above, upon studying functions \be{\cal L}_j(\alpha)=\prod_p 1/(1-\alpha^{\mid p\mid}L^{(p)}_j),\label{LFunc}\ee with $L^{(p)}_j$
defined in a similar manner as before,
for all integral  kernels $K_p$.

Furthermore, Taylor
expanding the scaling solution, indicative of the linear DSE, allows to parametrize it as a lower triangular matrix, which is a first sign of a connection to
integrable systems, which started to emerge in \cite{Kur2}.

What happens if our DSE is non-linear? This can be easily determined again under the condition that the non-linearity can be handled as in Eq.(\ref{der}) by
derivatives on the transform, using Fourier analysis and now a more general {\em Ansatz} $\exp{(-\gamma(\alpha)L)}\sum_j c_j(\alpha)L^j$. This reproduces the
propagator-coupling duality of \cite{BK}, this time originating from Fourier analysis. One uses the fact that any integral kernel provides an invariant operator
on the space of functions which has the transform defined above as an eigenvalue and $[q^2]^{-\gamma}$ necessarily as an eigenfunction.

The success in \cite{BK} the solution of a non-linear DSE is crucially based on the fact that the violations of scaling can still be expressed in terms of derivatives of
the transform of the underlying integral kernel. But this utterly fails when we have logarithmic corrections in different variables, correponding to different
loop momenta. The basic fact that the countertem which compensates a subdivergence does not store the information at which place a subdivergence is inserted allows
for a comparison of the problem with Galois theory. We finish this paper by a short first exposition of this analogy.

Let us take a quick glance at a second order algebraic equation $t^2+2a_1t+a_0=0$, with $a_1,a_0\in
\mathbb{Q}$, say. In general, there will be two roots $\rho_{1,2}= -a_1\pm\sqrt{a_1^2-a_0}$.
This defines two isomorphic extensions $\mathbb{Q}(\pm \sqrt{a_1^2-a_0})$, and the Galois group would be a
two element group here, consisting of the identity and the flip, realized as an automorphism
of the extended field for our normal extension. The equation itself remains invariant under the action of the Galois group.
Also, we note that the linear equation $x+a_0=0$ has trivially a solution in the field of coefficients: the solution to the linear equation
identifies the field to be extended by higher order equations.

Now consider $T[K_{p_1}]$ as above. It provides a solution to a linear DSE determining $\gamma(\alpha)=\sum_{k>0}c_k \alpha^k$.
For each order in $\alpha$, the coefficients $c_k$ are in some field, here $\mathbb{Q}$.

Now assume that we have non-trivial log insertions, which can be calculated from derivatives wrt $\gamma_1,\gamma_2$ on \be
T_2[K_{p_1}](\gamma_1,\gamma_2)\!:=\int\! d^6\!k \frac{ k^{-2\gamma_1} (k+q)^{-2\gamma_2} }{k^4\!(k+q)^2},\ee which determines the numbers generated from the
above primitive $p_1$ when log insertions are present for both internal propagators.
Comparing $T[K_{p_1}](\gamma_1+\gamma_2)$ with $ T_2[K_{p_1}](\gamma_1,\gamma_2)$, one immediately confirms the following facts:\\
i) there appear extensions $\mathbb{Q}(\zeta(m))$,\\
ii)  the same fields are generated when we exchange $\gamma_1$ and $\gamma_2$, only the rational coefficients change,\\
iii) $T[K_{p_1}](\gamma_1+\gamma_2)$ provides the fixed field invariant under the Galois group,\\
iv) $T_2[K_{p_1}](\gamma_1,\gamma_2)-T_2[K_{p_1}](\gamma_2,\gamma_1)$ plays the role of the discriminant: it provides the generator
of the extension and changes sign under the flip.

This interplay between transcendental extensions and finite group theory provided by the skeleton graphs, and its relation to Galois theory will be a crucial
ingredient yet to be developed to further our understanding of DSEs, with the two-loop master function \cite{Bierenbaum:2003ud} to be the next laboratory for
further analysis.

\end{document}